# Wavelength-dependent anisotropic light-matter interaction in 2D ferroelectric In$_2$Se$_3$

*Divya Jangra[1], Binoy Krishna De[1,2], Pragati Sharma[1], Koushik Chakraborty[1], Shubham Parate[2], Arvind Kumar Yogi[1], Ranjan Mittal[3,4], Mayanak K Gupta[3,4], Pavan Nukala[2], Praveen Kumar Velpula[1]\* and Vasant G. Sathe[1]\**
\*praveen@csr.res.in, vasant@csr.res.in

[1]UGC-DAE Consortium for Scientific Research, D.A. University Campus, Khandwa Road, Indore-452001, India

[2]Centre for Nano Science and Engineering, Indian Institute of Science, Bangalore-560012, India

[3]Solid State Physics Division, Bhabha Atomic Research Centre, Mumbai 400085, India

[4]Homi Bhabha National Institute, Anushaktinagar, Mumbai 400094, India

**Abstract**

The anisotropic light-matter interactions in 2D materials have garnered significant attention for their potential to develop futuristic polarization-based optoelectronic devices, such as photodetectors and photoactuators. In this study, we investigate the polarization-dependent interactions in ferroelectric 3R α-In$_2$Se$_3$ using Angle-Resolved Polarized Raman Spectroscopy (ARPRS) with different excitation lasers. Our experimental findings supported by complementary Density Functional Theory calculations demonstrate that the light-matter interactions depend not only on the crystallographic orientation but also on the excitation energy. Scanning transmission electron microscopy (STEM) confirms the highly anisotropic 3R crystal structure of α-In$_2$Se$_3$. This anisotropy in crystal structure facilitates significant optical anisotropy, driven by a complex interplay of electron-photon and electron-phonon interactions, which is reflected in the complex nature of the Raman tensor elements. Additionally, these anisotropy interactions extend to the material's electrical response under light illumination. Remarkably, the anisotropic photoresponse can be tuned by both polarization and wavelength of the incident light, making In$_2$Se$_3$ a promising material for advanced polarization-sensitive photodetection applications.

## 1 Introduction

The development of next-generation smart optoelectronic devices necessitates the exploration of new materials and innovative mechanisms. Traditionally, optoelectronic devices have relied on intensity-based signal detection and responsivity. However, recent trends have shifted towards harnessing sensitivity to light polarization, opening up new avenues for controlling conductivity [1], ferroelectric polarization [2], and dynamics of domains and domain walls [3,4,5] using polarized light. Two-dimensional (2D) materials have emerged as promising candidates for such studies due to their intrinsic anisotropy in crystal structure and electronic band distributions. These materials exhibit a wide range of unique optical, electronic, and magnetic properties, including optical anisotropy [6,7], flexible optical band structure [8,9,10], ferroelectricity [11,12,13], multiferrocity [14,15] and multifield-tunable light emissions [16,17]. These characteristics make 2D materials particularly ideal for polarization-sensitive device applications. As the demand for advanced functionalities in optoelectronics grows, research into the optoelectronic properties of 2D materials is evolving beyond conventional light intensity-based detection methods. This includes the investigation of the anisotropic photo response [18,19] and valley-polarization photoluminescence [20,21], achieved through the manipulation of incident light polarization. Understanding the polarization-dependent light-matter interaction in 2D materials is therefore critical to unravelling the underlying mechanisms governing these interactions. This understanding is essential for exploring the potential of 2D materials in future polarization-sensitive optoelectronic applications.

Angle-resolved polarized Raman spectroscopy (ARPRS) is a valuable non-destructive tool for studying the anisotropic light-matter interactions in 2D materials. It has been widely employed to explore the optical anisotropy in a range of materials, including $Bi_2Se_3$ [22], $PdSe_2$ [23], $MoS_2$ [24], $MoTe_2$ [25], black phosphorus [26,27], $ReSe_2$ [28,29] and $WSe_2$ [30]. The anisotropic crystal structure of 2D materials leads to variations in electron density and band distribution along different crystallographic directions, which in turn affects their electron-photon interactions. When a linearly polarized light interacts with these materials, the electron-photon interaction is enhanced if the polarization and energy of the photons align with favorable inter-band transitions [31]. Furthermore, the structural anisotropy of 2D materials is also evident in the behavior of lattice vibrations (phonons) along different crystallographic axes, resulting in anisotropic electron-phonon interactions [26]. Therefore, the overall light-matter interaction is governed by the combined contributions of both electron-photon and electron-phonon interactions. These interactions are highly sensitive to the energy and polarization state of the incident light. The anisotropic electron excitation (through electron-photon interaction) and subsequent electron-phonon coupling can lead to anisotropic photo-resistivity, offering a novel pathway to control the material's resistance by tuning the polarization of light.

$In_2Se_3$, a 2D ferroelectric, has garnered significant research interest due to its intriguing physical properties. These include room temperature in-plane and out-of-plane ferroelectric polarization, even down to monolayer thickness, as well as it's semiconducting nature, ease of integration with other 2D materials, and device stability [32]. Notably, it's in-plane and out-of-plane ferroelectric polarizations are inherently coupled, providing a unique opportunity to control the in-plane ferroic order parameter by applying an electric field along the out-of-plane direction [33,34,35]. Additionally, the persistence of the ferroelectric polarization in monolayer thickness makes it promising for high-density non-volatile memory devices, nano-electronics, field effect transistors, solar energy harvesting and sensors applications etc [36,37,38,39,40,41].In addition to ferroelectric properties, $In_2Se_3$ exhibits anisotropic crystal structure, along with a band gap energy~ 1.47 eV which can be tuned by layer thickness, make it suitable for photonics applications such as photodetectors and photo sensors [32]. Recent studies have demonstrated that the light polarization-dependent photoconductivity changes in nanoflakes of $2H-In_2Se_3$, which have been attributed to zig-zag and armchair atomic arrangements [42]. However, the specific atomic arrangements in $In_2Se_3$ remain to be fully established, and the microscopic mechanisms underlying it's light-matter interactions and anisotropic behavior still require further investigation.

Using the Angle-Resolved Polarized Raman Spectroscopy (ARPRS) with different excitation energies along with density functional theory (DFT) calculations, we investigated the light-matter interaction in α- $In_2Se_3$. Our results reveal that the material's highly anisotropic crystal structure plays a key role in driving its optical anisotropy, arising from a complex interplay of electron-photon and electron-phonon interactions. This interplay allows further tuning of the material's photoresponse by light polarization and wavelength, offering promising potential for applications in light polarization-based information detection and processing.

## 2 Results

## 2.1 Characterization of α-In₂Se₃ Single crystal

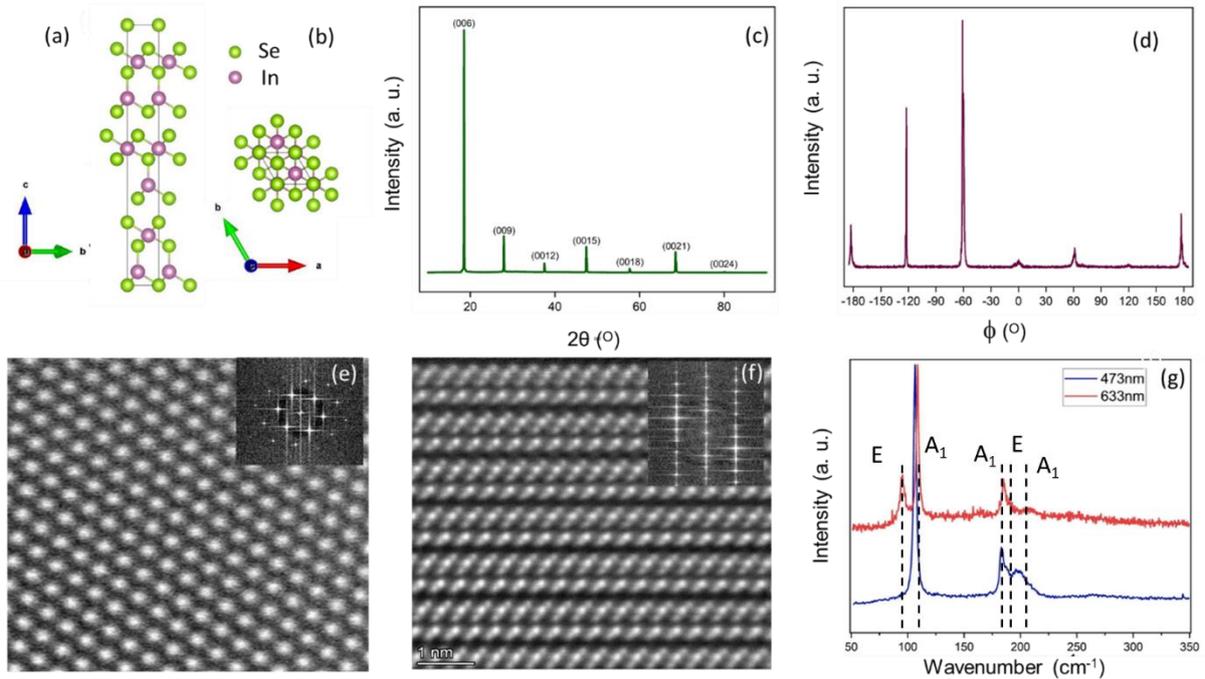

**Fig 1:** Characterization of In₂Se₃ Single Crystal. (a & b) Side and top-view of the crystal structure obtained using VESTA software. (c & d) X-Ray diffraction θ-2θ and phi-scan, respectively (e & f) HAADF-STEM image with *c* and *a* as the zone axes respectively. Inset shows the FFT pattern of the image of *ab* and *bc* plane (g) Raman spectra in *ab* plane using 473 nm and 633nm excitation laser.

The grown single crystal of In₂Se₃ was characterized using the X-ray diffraction (XRD), scaning transmission electron microscopy (STEM) and Raman spectroscopy studies. The crystal structure of α-In₂Se₃ belongs to rhombohedral crystal structure with the R3c space group [48]. Fig 1(a&b) represent the side (*bc* plane) and top (*ab* plane) views of the crystallographic unit cell, respectively, obtained using VESTA software [43]. The XRD pattern shown in Fig 1(c) confirmed that the single crystal is grown along (00C) direction and Phi scan on (104) reflection confirms the in-plane crystallinity as shown in Fig 1(d). The appearance of well-defined six peaks in phi scan corresponding to the (104) reflection confirms the rhombohedral structure of the crystal. Notably, α-In₂Se₃ can belong to 2H or 3R crystal structure depending upon the stacking sequence of the Se-In-Se-In-Se quintuple layers. The exact structure was clarified using aberration-corrected STEM. Atomically resolved high-angle annular dark field STEM (HAADF-STEM) images were acquired with zone axes as *c*-axis (Fig 1e) and *a*-axis (Fig 1f) confirming the 3R structure with a stacking period of three quintuple layers. Each quintuple layer consists of a sequential staking of Se-In-Se-In-Se layers seperated by approximately 2Å. Among the two In atoms in a quintuple layer, one In atom is at the center of an octahedral coordination provided by Se atoms and the other one lies at the center of the tetrahedral coordination by Se atoms. This structural arrangement is responsible for the ferroelectricity [44]. Crystal structure was further investigated by Raman spectroscopy measurements. The Raman spectra were taken using two different excitation wavelengths, 473nm and 633nm as shown in Fig 1(g). The blue and red Raman spectra correspond to 473nm and 633nm excitation lasers, respectively. The Raman spectra matched well with the previously reported Raman spectra of $\alpha - In_2Se_3$ [45]. Raman modes details will be discussed later. In earlier reports, Raman spectra were used to distinguish between 3R and 2H phases of α-In₂Se₃.

It was reported that the low energy E-mode is absent in case of 3R structure [45,46]. However, the present study shows that the presence/absence of low-energy E-mode (91cm$^{-1}$) depends on the excitation energy. This mode is observed when 633 nm excitation laser is used while absent when a 473 nm excitation laser is used, which may be due to the different electronic intermediate states are involved in the Raman scattering. [47]

The crystal structure is highly anisotropic in *bc* plane in contrast to the isotropic nature in *ab* plane. In this work, we explore how this anisotropic crystal structure influences the light-matter interactions and optoelectronic properties. To investigate the light-matter interaction and its dependence on photon energy, we performed ARPRS and photoresistivity measurements on both the *ab* and *bc* planes using 473 and 633nm excitation lasers as a function of light polarization angle θ.

## 2.2 ARPRS studies using 473nm and 633nm excitation laser

The 3R phase of α-In$_2$Se$_3$ belongs to the C$_{3v}$ point group and it's lattice vibrations can be written by the irreducible representation $\Gamma = 5A_1(R, IR) + 5E(R, IR)$, where A$_1$+E are acoustic modes, IR and R indicate the infrared and Raman active modes, respectively [48]. The experimentally obtained Raman modes were compared with those obtained by DFT calculations in Table 1. To investigate the energy and polarizaton dependence of light interaction in In$_2$Se$_3$, we performed ARPRS on *ab* and *bc* planes using linearly polarized 473 and 633 nm excitation lasers incident along *Z*-axis. A schematic of the ARPRS experimental geometries is shown in Fig 2(a,b). The ARPRS data were collected in two configurations. In the first configuration, the incident light propagation vector ($k_i$) was perpendicular to the *c*-axis of the sample, with the light polarization in the *bc* plane. An analyser was used to collect the Raman spectra such that the polarization of the incident ($e_i$) and scattered light ($e_s$) were parallel. The angle between the incident light polarization ($e_i$) and the crystallographic *b*-axis was varied by azimuthal rotation, denoted as rotation angle θ. This polarization geometry corresponds to a $a(bb)\bar{a}$ polarization configuration in Porto notation [49]. In the second configuration, the incident light wavevector ($k_i$) was parallel to the *c*-axis, with the light polarization in the *ab* plane of In$_2$Se$_3$. This configuration corresponds to $c(bb)\bar{c}$.

The contour plot of the Raman spectra as a function of θ for both excitation lasers, collected on the *bc* and *ab* planes are shown in Fig 2(c-f), with detailed spectra in Fig. S1 and S2. The Raman Modes were identified as follows: mode at 91cm$^{-1}$ corresponds to E mode, 104 cm$^{-1}$ mode corresponds to A$_1$(LO+TO) mode, 182 cm$^{-1}$ mode to A$_1$(LO) mode, 187 cm$^{-1}$ mode to E mode and 195 cm$^{-1}$ mode to A$_1$(LO), all these assigned modes were matched well with previous literature and our theoretical calculations (see Table 1) [48].

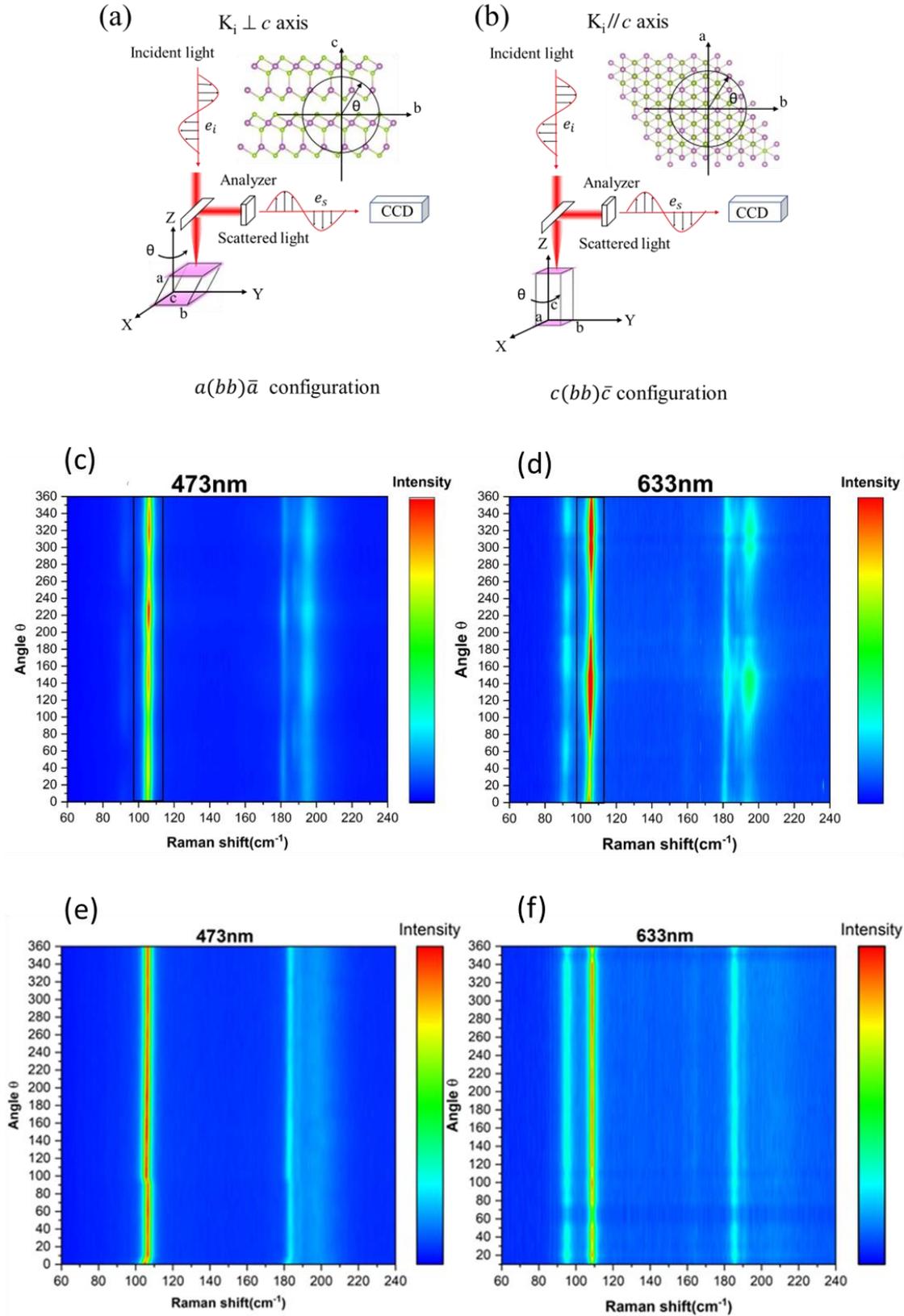

**Fig 2:** ARPRS results on *bc* and *ab* plane of In$_2$Se$_3$ single crystal are presented using two excitation laser sources 473nm and 633nm. (**a & b**) show the schematics of the experimental geometry in $a(bb)\bar{a}$ and $c(bb)\bar{c}$ polarization configuration in which the incident laser is perpendicular to *bc*- and *ab*-plane, respectively. XYZ and abc are the laboratory and crystallographic unit cell axis, respectively. θ is the angle between light polarization direction and crystallographic *b*-axis. $e_i$ and $e_s$ are the polarization direction of the incident and scattered light. Pink shaded area represents the crystallographic planes. (**c & d**) contour plot of Raman spectra as a function of θ collected on *bc* plane using 473nm and 633nm excitation source respectively. (**e & f**) contour plot of Raman spectra as a function of θ collected on *ab* plane using 473nm and 633nm excitation source respectively.

**Table 1:** Raman modes frequency and tensor elements for different excitation energies along with theoretically calculated values.

| mode | ω theoretical (cm-1) | ω exp (cm-1) | Theoretical a/b | Theoretical c/d | 473nm a/b | 473nm c/d | 473nm Phase (°) | 633nm a/b | 633nm c/d | 633nm Phase (°) |
|---|---|---|---|---|---|---|---|---|---|---|
| E(1) | 24.8529 | | | | | | | | | |
| E(2) | 94.3524 | 91 | | 0.29 | | 1.13 | 102 | | 0.2 | 0 |
| A$_1$(1) | 105.354 | 104 | 1.85 | | 1.01 | | 52 | 1.52 | | 0 |
| E(3) | 155.664 | | | | | | | | | |
| A$_1$(2) | 182.46 | 182 | 0.42 | | .95 | | 58 | 1.41 | | 48 |
| E(4) | 183.32 | 187 | | 0.94 | | .39 | 115 | | 0.48 | 83 |
| A$_1$(3) | 196.755 | 195 | 1.28 | | 1.04 | | 47 | 1.48 | | 49 |
| A$_1$(4) | 249.994 | | | | | | | | | |

Notably, the E(2) mode was appeared only in spectra collected using 633nm excitation laser, which can be attributed to different electronic intermediate states are involved in the Raman scattering [30,47,50]. The angular dependence of the Raman modes intensity in the *ab* plane shows invariance with respect to θ (Fig S3(a)), indicating isotropic light scattering and no polarization dependence. In contrast, the *bc* plane Raman spectra show a strong dependence on the light polarization angle. To get the quantitative information of intensity variation with θ, the Raman spectra were fitted using Lorentz function and obtained intensities were plotted as a function of azimuthal angle θ in Fig 3 and Fig S3(b). Green dots represent the experimental data. The intensity of all the Raman modes changes with the light polarization angle, highlighting the anisotropic nature of the light-matter interaction in the *bc* plane, in contrast to the isotropic nature in *ab* plane. Notably, the A$_1$(1) and E(2) modes exhibit different angular variations for 473 nm and 633 nm excitations (Fig 3). The A$_1$(1) mode displays a 90° periodic variation with 473nm laser and a 180° variation with 633nm laser. These results suggest that the light-mater interaction in the *bc* plane is anisotropic in nature and depends on the energy of the incident light. To further understand this anisotropy, we simulated the intensity variation with respect to the incident light polarization angle using Raman tensors.

The polarized Raman intensity can be expressed as [6,10,19,22]

$$I = (\langle e_i|\mathbf{R}|e_s\rangle)^2 \qquad (1)$$

where $e_i = (0 \quad \cos\theta \quad \sin\theta)$ and $e_s = (0 \quad \cos\theta \quad \sin\theta)'$ represent the polarization directions of incident and scattered light and R is the Raman tensor. The Raman tensors for the $A_1$ and E modes for $C_{3v}$ symmetry are as follows:

$$\mathbf{R}(A_1) = \begin{bmatrix} ae^{i\phi_a} & 0 & 0 \\ 0 & ae^{i\phi_a} & 0 \\ 0 & 0 & be^{i\phi_b} \end{bmatrix} \tag{2}$$

For E modes are double degenerate mode and the two possible forms of $\mathbf{R}(E)$ are as follows:

$$\mathbf{R}(E) = \begin{bmatrix} 0 & ce^{i\phi_c} & de^{i\phi_d} \\ ce^{i\phi_c} & 0 & 0 \\ de^{i\phi_d} & 0 & 0 \end{bmatrix} \text{ or } \mathbf{R}(E) = \begin{bmatrix} ce^{i\phi_c} & 0 & 0 \\ 0 & ce^{i\phi_c} & de^{i\phi_d} \\ 0 & de^{i\phi_d} & 0 \end{bmatrix} \tag{3}$$

Where $a$, $b$, $c$, and $d$ are Raman tensor elements and $\phi_a, \phi_b, \phi_c$ and $\phi_d$ are the phases of the corresponding Raman tensor elements. Substituting the Raman tensor expressions in equation (1), the intensity equations for $bc$ and $ab$ planes as a function of the crystal orientation $\theta$ in the parallel polarization configuration are as follows:

For the $A_1$ mode in $bc$ plane:

$$I_A(bc) = |a|^2 Sin^4(\theta) + |b|^2 Cos^4(\theta) + (|a||b|Sin^2(2\theta)Cos(\phi_{ab}))/2 \tag{4}$$

For the E mode in $bc$ plane:

$$I_E(bc) = |c|^2 Cos^4(\theta) + |d|^2 Sin^2(2\theta) - 2|c||d|Sin(2\theta)Cos^2(\theta)Cos(\phi_{cd}) \tag{5}$$

For the $A_1$ mode in $ab$ plane: $I_A(ab) = |a|^2$ \hfill (6)

For the E mode in $ab$ plane: $I_E(ab) = |c|^2$ \hfill (7)

Thus, the intensity of the $A_1$ and E modes depends on the magnitude of the Raman tensor elements, the relative phase $\phi_{ab} = \phi_a - \phi_b$ and azimuthal rotation angle $\theta$. The Raman tensor elements are functions of the dielectric susceptibility and are generally complex, introducing a phase factor. This complex nature becomes particularly significant in cases involving optical absorption i.e. when the excitation photon energy ($E_{ex}$) is equal to or greater than the electronic excitation energy gap ($E_g$). In the present case, both the used excitation laser energies (1.96 eV and 2.6 eV) are higher than the bandgap energy (1.4eV, calculated using a Tauc plot; see Fig S9). It is important to note that we could not determine the absolute phase of a Raman tensor element because the intensity is proportional to the square of the tensor elements, meaning only the relative phase can be extracted by fitting the experimental intensity to the theoretically obtained equations. In general, when $E_{ex} < E_g$, the Raman tensor can be considered to be real, implying that the phase difference should be zero. Using the theoretical equations, we attempted to fit the angular- dependent Raman intensity by considering real Raman tensor elements i.e. by considering relative phase difference of zero (see Note S1). The Raman mode $A_1(1)$ fit well for 633nm laser, but did not fit well for 473nm laser. The experimental intensity as a function of light polarization angle $\theta$ has been fitted with (solid line) and without (dash line) considering phase (See Fig 3). Similarly, we fitted the intensity of the E(2) mode and found that it fitted well when assuming Raman tensor elements are real for the 633 nm Laser. On the other hand, introducing a phase factor was necessary to properly fit the intensity data for the 473 nm laser, indicating that the Raman tensor elements exhibit a complex nature at this excitation wavelength. Other Raman modes such as $A_1(2)$, $E(4)$, and $A_1(3)$ can be fitted only by considering the complex nature of Raman tensor elements for both the laser excitations See in supplementary (Fig S3(b)).

The experimentally determined values of the Raman tensor elements depend on factors such as laser intensity, experimental geometry, and detector quantum efficiency, rather than their absolute values alone. To correct these experimental variables, we compared the ratio of Raman tensor elements. The experimentally obtained ratios of the Raman tensor elements for $A_1$ (a/b) and E (c/d) modes, along with their phase difference, are listed in Table 1. Notably, the phase difference was found to be larger for the 473nm laser excitation. Interestingly, the Raman tensor elements for both excitation lasers were found to be real and isotropic in the *ab* plane. These results indicate that the polarized Raman intensity depends on both the light polarization and excitation wavelength.

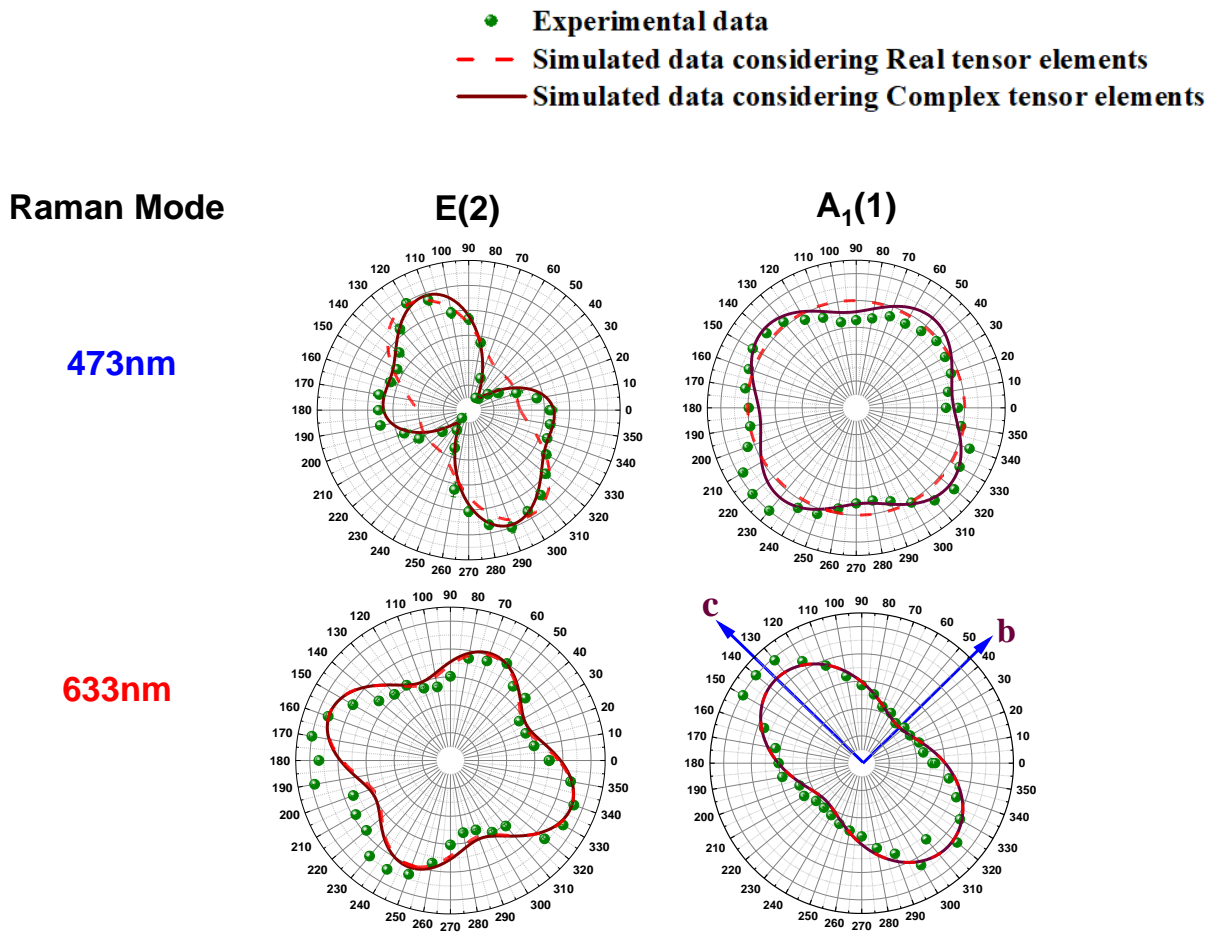

**Fig 3**: The intensity variation of the E(2) and $A_1(1)$ mode as a function of incident light polarization angle or azimuthal angle $\theta$, relative to the *b*-axis of the crystal in the *bc* plane was measured using 473 nm and 633 nm excitation lasers. Blue arrows indicate the crystallographic directions.

## 2.3 Theoretical calculations

We performed phonon calculations based on density functional theory (DFT) to analyse and interpret the phonon modes observed in Raman spectra and determine their vibrational nature. The calculated phonon wavenumbers at the zone center matched well with the experimentally observed phonon wavenumbers (listed in Table 1). To further understand the anisotropy and its effect on the Raman mode intensity in different planes, we calculated the Raman tensor of all the active modes. A comparison of the DFT calculated and experimentally obtained ratios of the Raman tensor elements is provided in Table 1. In our theoretical calculations, we did not explicitly consider the electron-phonon interaction term, which corresponds to phonon-induced electronic transitions. Consequently, the calculation gives only the real tensor elements, which closely match with the results from the 633nm excitation laser, where the phase is negligible. However, the tensor element ratios for the 473 nm laser differ significantly between the calculated and experimental values, with the phase of the tensor elements being much larger. This suggests strong electron-phonon/photon interaction originated from the exciton-phonon interaction, which will be discussed later. Our results indicate that the Raman tensor phase value is not solely determined by the intrinsic properties of the Raman tensor but also depends on the energy of the excitation laser used in the measurements.

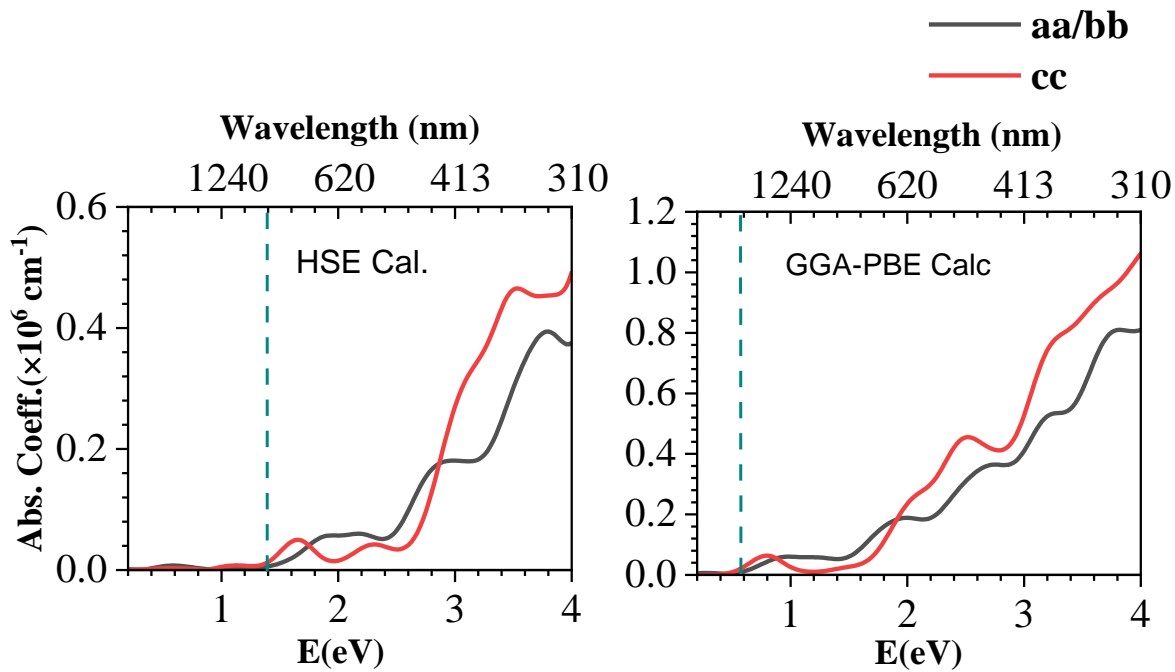

**Fig 4**: Theoretically calculated absorption spectra using Hybrid exchange correlation functional (HSE) and GGA-PBE exchange correlation functional along different crystallographic axes. Black and red lines represent absorption along *c* and *a/b* axis.

In order to understand the role of electron-photon interaction, we calculated absorption coefficients along crystallographic axes (*a*, *b*, and *c*) using Hybrid exchange correlation functional (HSE) and GGA-PBE exchange correlation functional shown in Fig 4. The optical absorption was found to be almost isotropic for incident photon energy 1.96 eV (633nm) while significantly higher absorption along the *c*-axis for 2.62 eV (473 nm) laser. This isotropic absorption for 633nm along the three crystallographic axes indicates the isotropic electron-photon interaction whereas anisotopic interaction for 473 nm. These results indicate that the anisotropy in absorption i.e., electron-photon interaction linked to the incident photon energy.

From the above findings, it is evident that the polarized Raman intensity exhibits significant anisotropy in the *bc* plane. It may be noted that the symmetry of the eigen-displacement of a phonon mode plays a crucial role in driving this anisotropic Raman intensity variation [26,27]. To explore this, we calculated the eigen displacements using the Phonopy software. Eigen displacements are shown in Fig 6 and Fig S4. The results indicate that the singly degenerate $A_1$ modes correspond to atomic vibrations along *c*-axis (out-of-plane vibration), while the doubly degenerate E modes exhibit in-plane atomic vibrations within the *ab* plane. The contribution of the phonon symmetry to the observed Raman intensity anisotropy is further discussed in the discussion section.

## 2.5 Polarization-dependent optical reflectivity study

We conducted angle-resolved optical reflectivity measurements using 488 nm and 633 nm lasers to explore anisotropy in optical absorption. Details of the measurement are discussed in the experimental section. The optical setup, shown in Figure S11(a), used back-illuminated geometry and parallel polarization configuration to capture angle-resolved reflectivity contrast images. Fig. S11(b,c) shows the relative contrast of the reflected intensities as a function of the light polarization angle. Our observations revealed that the relative contrast in reflectivity is anisotropic for 488 nm, with minimum contrast along the c-axis. This suggests higher absorption along the c-axis, consistent with theoretical measurements using 473 nm. In contrast, the reflectivity contrast for 633 nm is nearly isotropic, aligning with theoretical isotropic absorption using 633 nm. (Details of the measurement are discussed in the experimental section). In the zero transmission limit, absorption is complementary to the reflectivity of the material. That means, higher reflectivity for a particular light wavelength implies lower absorption at that wavelength, and vice versa. It is observed that the optical absorption (complementary of the reflectivity as shown in fig S11(b)) is highly anisotropic for 488 nm, it is maximum along *c* axis. While the absorption is isotropic for 633nm excitation laser (fig S11 (c)).

## 2.6 Angle-resolved photoresistivity measurements

The anisotropic light-matter interaction arises due to polarization-dependent electron-phonon-photon interaction, which can results in a polarization-dependent photoconductivity. Due to anisotropic optical absorption, the photocurrent can show a periodic variation under the application of polarized light [51]. To understand the effect of anisotropic light-matter interaction on photoconductivity, we performed photoconductivity measurements as a function of polarization rotation using 473 and 633nm lasers. Experimental details are in note S2. Fig S6(a,b,c&d) shows the photoconductivity switching by light on *bc* and *ab* plane using 473nm and 633nm, respectively. Further, to rule out the photothermal effect-induced resistive switching, we have performed the resistive switching as a function of incident laser power and the results are shown in Fig S7.

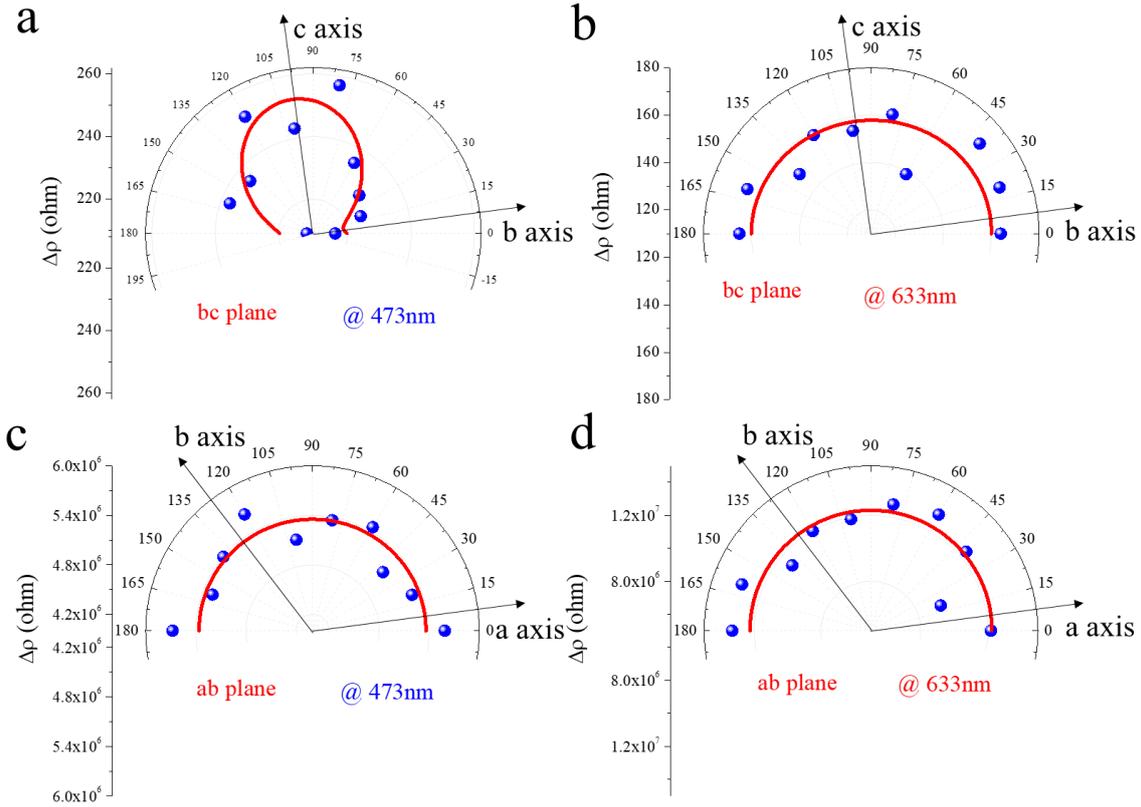

**Fig 5:** Photoresistivity as a function of light polarization angle. (a & b) and (c & d) photo resistivity in *bc* and *ab* plane under 473 nm and 633 nm, respectively. Anisotropic photo response is observed when light of wave length 473nm is incident on the *bc* plane, whereas photocconducctivity is isotropy for 633nm laser. When measured on *ab* plane the photocconducctivity is isotropiy for both the laser excitation wavelength.

The quantative photoresistive switching in *bc* and *ab* plane as a function of light polarization is shown in Fig 5(a-d). For 633 nm laser excitation, the photoresponse is almost isotropic in both *ab* and *bc* plane while for 473nm laser excitation the response in *bc* plane is anisotropic and shows dependence on the incident light polarization angle (Fig 5(a)). For 473nm, the change in resistivity is maximum along *the c-axis* and minimum along the *b* direction, and the anisotropy is ~ 15.5 %  (see Note S3).

## 3 Discussion

The quantum picture of light-matter interaction offers deeper insight into the observed anisotropy. In quantum mechanics, Raman scattering is described as a third-order time-dependent perturbation process and the intensity of $\nu^{th}$ mode can be expressed as [52]

$$I^\nu(E_L) = \left| \sum_{i,m,m'} \frac{\langle f|H_{op}|m'\rangle \; \langle m'|H_{e-ph}|m\rangle \; \langle m|H_{op}|i\rangle}{(E_L - \Delta E_{m'i}) \; (E_L - \hbar\omega_\nu - \Delta E_{mi})} \right|^2 \qquad (8)$$

Where $H_{op}$ and $H_{e-ph}$ are the Hamiltonian for optical transition and electron-phonon interaction, respectively, $\Delta E_{m'i} = E_{m'} - E_i - i\Gamma_{m'}$ and $\Delta E_{mi} = E_m - E_i - i\Gamma_m$ are the energy differences between m'-th and m-th to i-th energy levels and $\Gamma$ - is the broadening factor related to the lifetimes of the excited states involved.

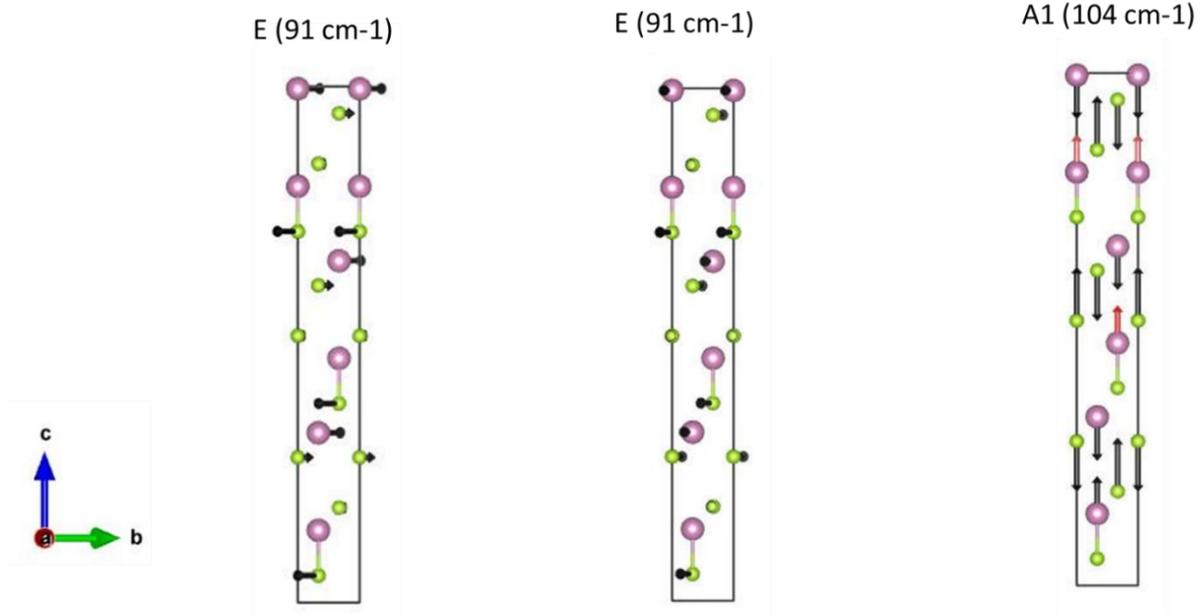

**Fig 6:** Eigen displacement of $A_1(1)$ and $E(2)$ Raman modes. The two possible Eigen displacements of the duble dedgenerate E mode is shown in first two image from left.

The Raman scattering process involves three processes: (1) absorption of a photon, exciting an electron from the initial state *i* to the state *m* through electron-photon interaction $H_{op}$. (2) The excited electron interacts with the phonon ($H_{e-ph}$) and goes to another excited state m'. (3) The excited electron decays to the final state f with the emission of scattered radiation. So, the complete Raman scattering process involves two optical transitions and an electron-phonon interaction. Other sequences of these processes are also possible [52]. According to equation 8, the Raman tensors depend on the excitation energy, phonon energy, and interplay of electron-photon and electron-phonon interactions. Thus, the effective light-matter interaction is influenced by the symmetries of both the electronic states (through electron-photon interaction) and phonon states (through electron-phonon interaction) involved.

As we have seen from the optical reflectivity studies and theoretical calculations, isotropic absorption for 633 nm along the three crystallographic axes indicates the isotropic electron-photon interaction. Therefore, the observed higher intensity of $A_1$ Raman modes along *c* axis (Fig 3) could be due to the electron-phonon interaction anisotropy originated from the phonon symmetry. The Eigen displacements of the $A_1$ Raman modes (shown in Fig 6 and S4) show atomic vibrations along *c*-axis, resulting in stronger electron-phonon interactions along *c*-axis. This, combined with the phonon symmetry results in a two-fold symmetry (180° lobes) in ARPRS intensity plots, which can be fitted by considering real Raman tensor elements. In contrast, the ARPRS intensity plots of the $A_1(1)$ mode obtained with 473nm excitation showed a four-fold symmetry, which could not be explained by electron-phonon interactions alone. This indicates a complex interplay between electron-phonon and electron-photon interactions. The anisotropic electron-phonon and electron-photon interaction introduce relative phase angles in Raman tensor elements, as shown in fitting results (Table 1). The relative phase angles for low-energy Raman modes $E(2)$ and $A_1(1)$ are larger for 473 nm laser excitation, indicating stronger coupling between these phonon modes and excited electronic state. Thus, we conclude that both electron-photon and electron-phonon interactions contribute to the anisotropic light-matter interaction observed in In$_2$Se$_3$, explaining the different anisotropic behaviour of the same mode under different excitation energies.

Scanning transmission electron microscopy (STEM) studies have shown that the crystal structure is isotropic in the *ab* plane but highly anisotropic in *bc* plane. The different atomic arrangement and interatomic interaction strength along the *c*-axis to its perpendicular axes produce anisotropic electronic distributions or bands as shown in Fig S8 (calculated band structure), leading to optical anisotropy through the complex interplay of electron-photon and electron-phonon interactions.

It's been reported that the anisotropic photo-responses can arise from the bulk photovoltaic effect and photoconductivity anisotropy [4]. However, our I-V characteristic measurements with and without light (see Fig S10) show negligible contribution of bulk photovoltaic effect [53] which is probably below our detection limit. Hence, we can neglect the bulk photovoltaic effect and conclude that the photo response anisotropy can depend only on the photoconductivity anisotropy arising due to exciton generation. Furthermore, the resistivity can change under light illumination due to photothermal effect [54]. The saturation of the photoresistivity in power-dependent photo-resistivity studies (shown in Fig S7(c,d)) rules out the possibility of photo thermal effect in the observed photo response [54,55] (details is in Note S4).

It is important to note that the photo response in 3R α –In$_2$Se$_3$ is significantly large. , It is 70% for 633 nm (maximum varied laser power up to 2.5 mW) and 35% for 473 nm (maximum varied laser power up to 3.5mW) laser illumination. The photoconductivity is defined as σ = nqμ, where it depends on photo-excited charge carrier density (n) and carrier mobility (μ) in the conduction band. The charge carrier density depends on the charge carrier generation and recombination which depends on electron-photon and electron-phonon interaction, and carrier mobility depends on the electron-phonon interaction [56]. Anisotropic electron-phonon interaction may produce directional-dependent conductivity. However, here we have done the photoconductivity measurements along *c* direction with varying light polarization angle, so the mobility (electron-phonon interaction) is same for all polarization dependent measurements. Therefore, only the electron-photon interaction (exciton generation) can introduce anisotropy in photo-conductivity.

For 473 nm excitation, our ARPRS studies indicate highly anisotropic electron-phonon and electron-photon interactions in the *bc* plane, while these interactions are isotropic in *ab* plane, leading to the observed anisotropic photo-response in the *bc* plane (Fig 5). The photo-induced conductivity enhancement is larger along *c*- axis, attributed to the increase in the optical absorption coefficient along *c*-axis for 473 nm laser excitation, resulting in higher carrier generation. In contrast, for 633nm laser excitation, where electron–photon interactions are minimal (as reflected by the negligible phase angles of Raman tensor elements), the photo-response is isotropic (see Fig 5b and d). These results demonstrate that the photo-response in In$_2$Se$_3$ is both energy of the excitation- and crystallographic direction-dependent and can be tuned by light polarization.

## 4 Conclusion

In summary, our experimental measurements combined with DFT calculations, reveal pronounced anisotropy in the light-matter interaction within the *bc* plane of 3R α-In$_2$Se$_3$ single crystal, while the interaction remains isotropic in *ab* plane. This anisotropy in the *bc* plane arises from the distinct atomic arrangement along different crystallographic axes, leading to variations in electronic distribution and phonon symmetry. These factors contribute to the complex nature of Raman tensors, which are governed by a complex interplay between electron-photon and electron-phonon interactions. Consequently, the Raman tensor elements and their relative phases exhibit strong dependence on both the excitation energy and polarization state of incident light. The observed anisotropic light scattering in $In_2Se_3$ governed by these interactions resulting in an excitation-wavelength-dependent photoresponse. This ability to tune the photoresponse based on light polarization and photon energy underscores the potential of $In_2Se_3$ for futuristic applications in polarization-sensitive devices.

## 5 Experimental Technique

**Single crystal growth and XRD θ-2θ and phi scan**

High-quality single crystals of $In_2Se_3$ were grown from the ultra-high pure 5N constituent elements of In and Se directly via a modified Bridgman method by using an appropriate temperature profile along with a careful optimization of the modified Bridgman growth method. The growth was performed in sharp cone shaped one end of the quartz tube and the used dimensions ~16 mm diameter, 18 cm long quartz tube. The quartz tube was cleaned using HF/H2O (1:10) solution to removing impurities from the quartz tube and thoroughly rinsed with distilled water prior to the growth process. This process also makes the inner surface of quartz tube a slight rough which is beneficial in promoting nucleation during the crystal growth. The constituent elements were weighed in the stoichiometric ratio individually and then introduced into the quartz tube. The quartz tube with desired elements was sealed under a vacuum better than $2 \times 10^{-6}$ Torr. The growth was initiated by placing the quartz tube into a high temperature box furnace. The temperature was increased to 1075 °C slowly over a 72 h period and hold it for 24 h. The temperature was very slowly decreased to 700 °C at the rate of 1 °C/h over a period of about 15 days, after which the tube was cooled down to room temperature at the rate of 60 °C/h. This resulted in the formation of shiny ingot crystals approximately ~14 × 20 mm2 in lateral size, which is readily cleavable in layers of very thin platelike layers. Required crystals for various measurements were thin 2D layers of $In_2Se_3$ crystals, which were produced by micro-mechanical cleavage from the bulk shiny as grown ingot of $In_2Se_s$. To clarify the crystal structure, symmetry, crystal orientation, and phase of an as-grown single crystal of $In_2Se_3$, we have used HR-x-Ray diffraction and Raman spectroscopy measurements were taken. For the or $\theta - 2\theta$ scan we used Cu K-Line for diffraction measurement and Bruker D2 phaser diffractometer in Bragg-Brentano geometry. For $\phi$ scan measurement PAN analytical X`PERT high-resolution x-ray diffraction (HR-XRD) system equipped with a Cu anode was used.

**ARPRS measurements**

For Raman characterization, we used a Raman spectrometer from Horiba Jobin Yvon (France) HR-800 equipped with a diode laser 473nm and He-Ne laser 633nm, 1800 g mm$^{-1}$ grating, and a Peltier cooled CCD detector with a Spectral Resolution of ~1 cm$^{-1}$. Laser beam focused onto a sample with a spot size of ~1μm by a 50× lens with an exposure time of 60 sec. and 2 accumulations. The power of a laser is controlled by using Filters to avoid the local damage in the Sample due to laser heating. We used a 10% filter to avoid sample heating. Raman spectra were collected from the 50-350 cm$^{-1}$ range. For Angle Resolved Polarized Raman spectra, the Sample is mounted on a Rotating stage with a 0-360$^O$ angle marker. An analyser is used to collect the scattered radiation.

## Scanning transmission electron microscope

STEM images were collected using aberration (probe) corrected Titan Themis 300 operated at 300kV equipped with SuperXG quad EDS detector. 24 mrad convergence angle used for EDS and HAADF collection angle of 48-200 mrad were used at a camera length of 160mm.

## Photoconductivity Experimental Setup

Photoconductivity measurements are performed in the linear two point probe configuration using Keithley 2425 source meter and Keithley 2002 multimeter. Light on/off studies are done using 1.25 mW and 1.75 mW power of 473nm and 633nm laser, respectively. Resistivity measurements was done using 1μA.

## Optical reflectivity study

Polarized optical microscopy with angle-resolved analysis was used to examine the absorption anisotropy using 488 nm and 633 nm laser light. The optical setup, shown in Figure S11(a), used back-illuminated geometry and parallel polarization configuration to capture angle-resolved reflectivity contrast images. A CCD camera was used to acquire the reflectivity contrast images. We rotated the sample using a homemade rotating stage to change the light polarization with respect to the crystallographic axis. The optical image of the bc plane was analyzed using ImageJ software, and the reflected intensity contrast was determined from the brightness of a selected area of the image. All measurements involving polarization angle were carried out with a constant laser power. The exposure and image acquisition conditions were kept constant throughout the measurements.

## DFT calculation

First-principles DFT calculations were conducted using the Vienna Ab Initio Simulation Package (VASP). Valence electrons were represented by the projector-augmented wave (PAW) method. Plane-wave expansions used the default energy cutoffs provided by VASP's PAW potentials. The PBE parametrization of GGA was employed for exchange and correlation functionals in structural relaxations and total energy calculations. The HSE06 hybrid functional calculations were performed to precisely estimate the band structure, density of states, and optical absorption properties. A plane-wave energy cutoff of 700 eV was used in all calculations. A Γ-centered 12×12×12 Monkhorst-Pack k-mesh was used for k-point sampling. Atomic structures were optimized until forces on all atoms were less than 0.001 eV/Å. Van der Waals corrections were included in heterostructure calculations using the semiempirical DFT-D3 method."

## Acknowledgement

Authors acknowledge the Advanced Facility for Microscopy and Microanalysis (AFMM), Indian Institute of Science, Bengaluru, India for STEM measurements, and Micro nano characterization facility, CeNSE, IISc for FIB based sample preparation. P.K.V. acknowledges the financial support from the Science and Engineering Research Board (SERB) project SRG/2022/000118.

## Conflict of Interest

The authors declare no conflict of interest.

## Data Availability Statement

The data that support the findings of this study are available from the corresponding author upon reasonable request.

## Keywords

Optical anisotropy, Polarized Raman spectroscopy, Photo-response, Light-matter interaction, Electron-photon-phonon interaction, light polarization based optoelectronics.

# Supplementary Information for

## Wavelength-dependent anisotropic light-matter interaction in 2D ferroelectric In$_2$Se$_3$

*Divya Jangra[1], Binoy Krishna De[1,2], Pragati Sharma[1], Koushik Chakraborty[1], Shubham Parate[2], Arvind Kumar Yogi[1], Ranjan Mittal[3,4], Mayanak K Gupta[3,4], Pavan Nukala[2], Praveen Kumar Velpula[1]\* and Vasant G. Sathe[1]\**
\*praveen@csr.res.in, vasant@csr.res.in

[1]UGC-DAE Consortium for Scientific Research, D.A. University Campus, Khandwa Road, Indore-452001, India

[2]Centre for Nano Science and Engineering, Indian Institute of Science, Bangalore-560012, India

[3]Solid State Physics Division, Bhabha Atomic Research Centre, Mumbai 400085, India

[4]Homi Bhabha National Institute, Anushaktinagar, Mumbai 400094, India

**Table of content**

Note S1: Raman Intensity Equations Considering Real Tensor Elements

Note S2: Photoconductivity Experimental Setup

Note S3: anisotropic photo response calculation

Note S4: Power dependent study

Note S5: V-I characteristics under different light

**Angle dependent polarized Raman spectra in *bc* and *ab* plane using 473nm and 633nm excitation laser**

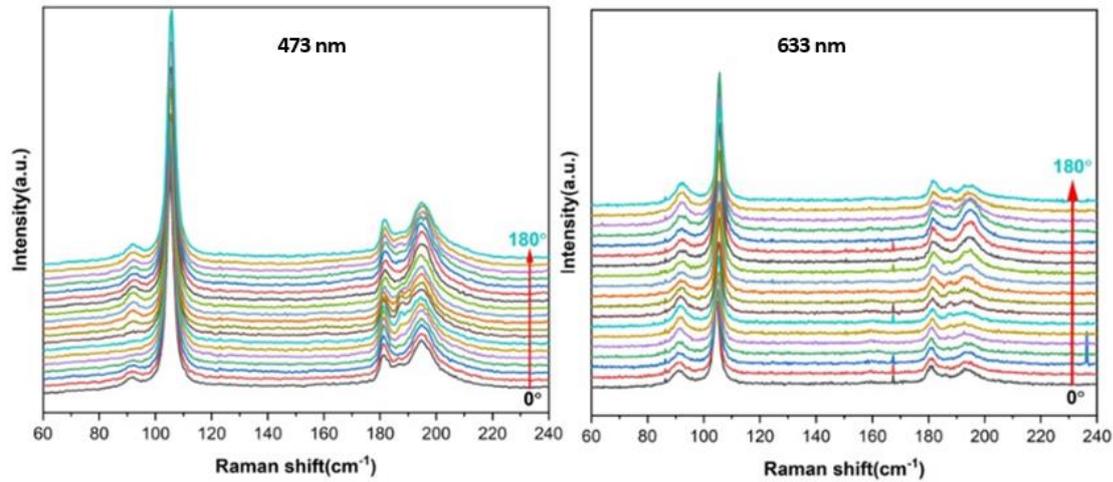

Fig S1: Angle Dependent Polarized Raman data in *bc* Plane Using 473 nm and 633 nm Excitation Source

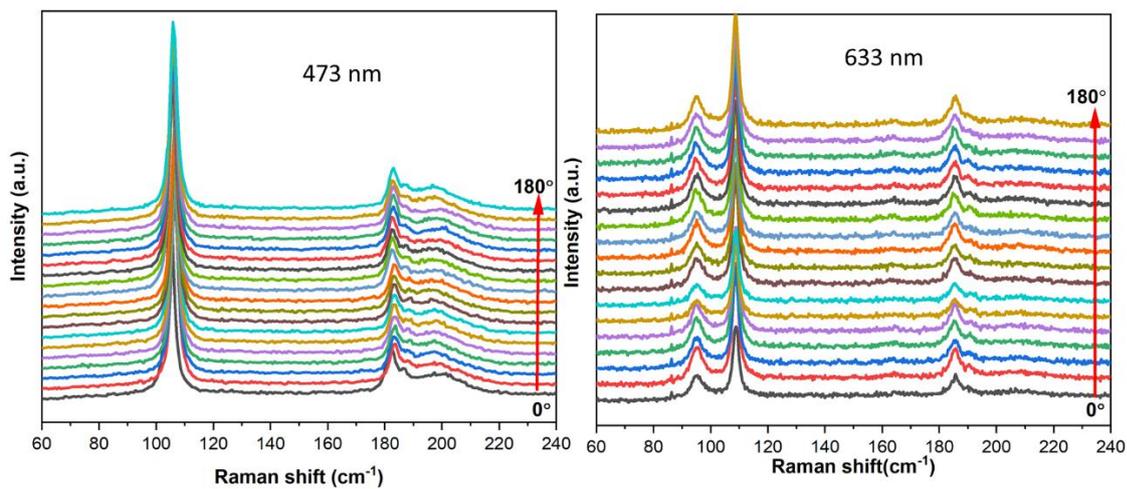

Fig S2: Angle Dependent Polarized Raman data in *ab* Plane using 473 nm and 633 nm Excitation Source.

**Note S1: Raman intensity equations considering real tensor elements**

Raman intensity equation for *bc* plane by considering real tensor element only by taking the cosine of phase difference to be unity

$$I_A(bc) = |a|^2 Sin^4(\theta) + |b|^2 Cos^4(\theta) + (|a||b|Sin^2(2\theta))/2$$

$$I_E(bc) = |c|^2 Cos^4(\theta) + |d|^2 Sin^2(2\theta) - 2|c||d|Sin(2\theta)Cos^2(\theta)$$

(a)

Experimental data
Simulated data

| Raman mode | E mode | A mode |
|---|---|---|
| 473nm | 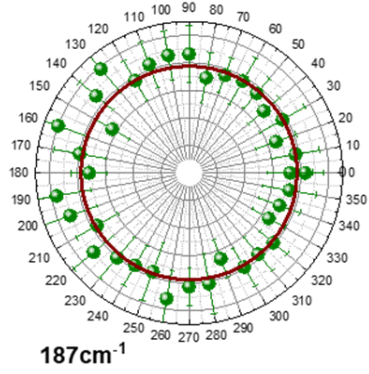 187cm$^{-1}$ | 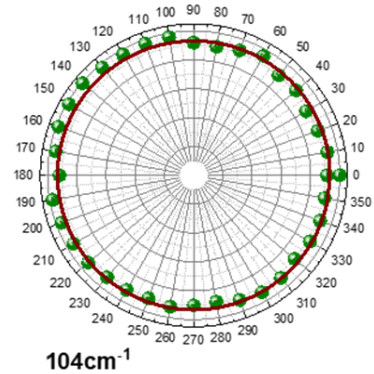 104cm$^{-1}$ |
| 633nm | 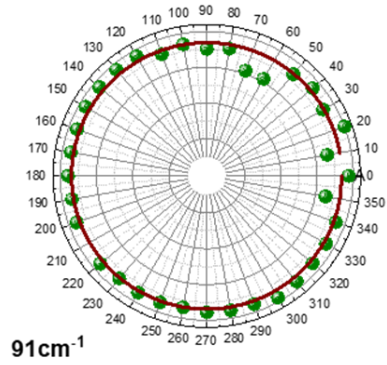 91cm$^{-1}$ | 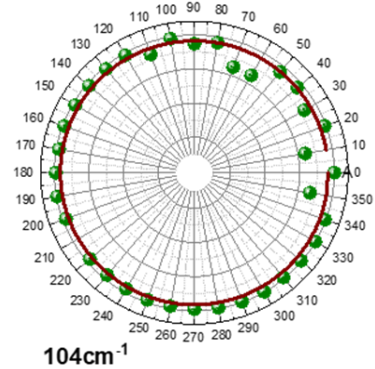 104cm$^{-1}$ |

(b)

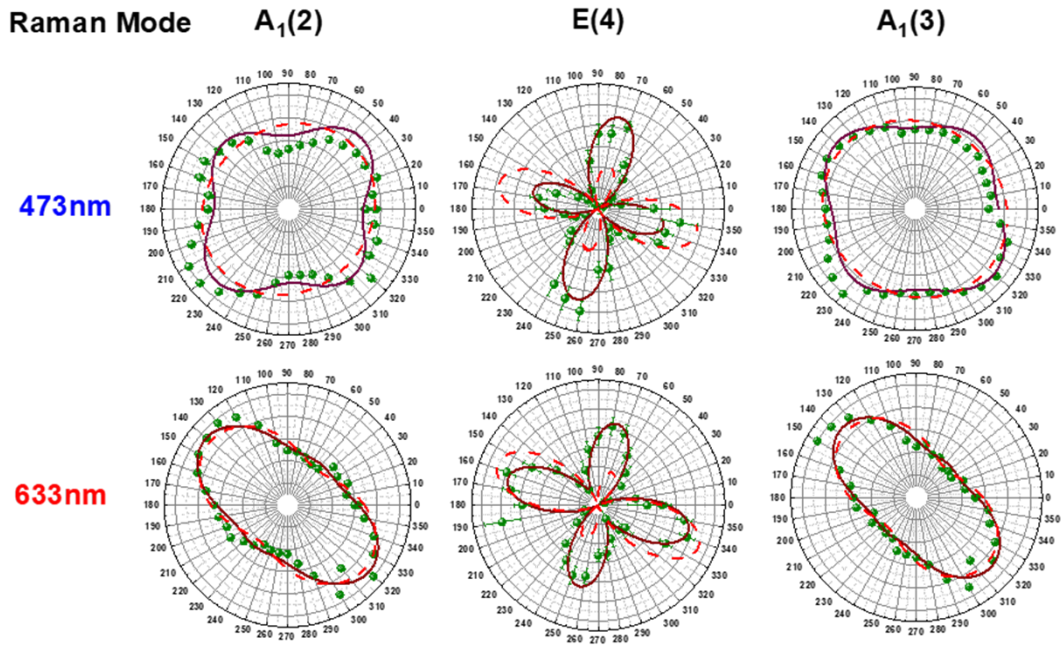

Fig S3: (a) Polar plot of $A_1$ and E mode in *ab* plane using both excitation source. (b) Polar plot of A and E mode in *bc* plane using both excitation source.

**Theoretically calculated Eigen displacement of A and E modes**

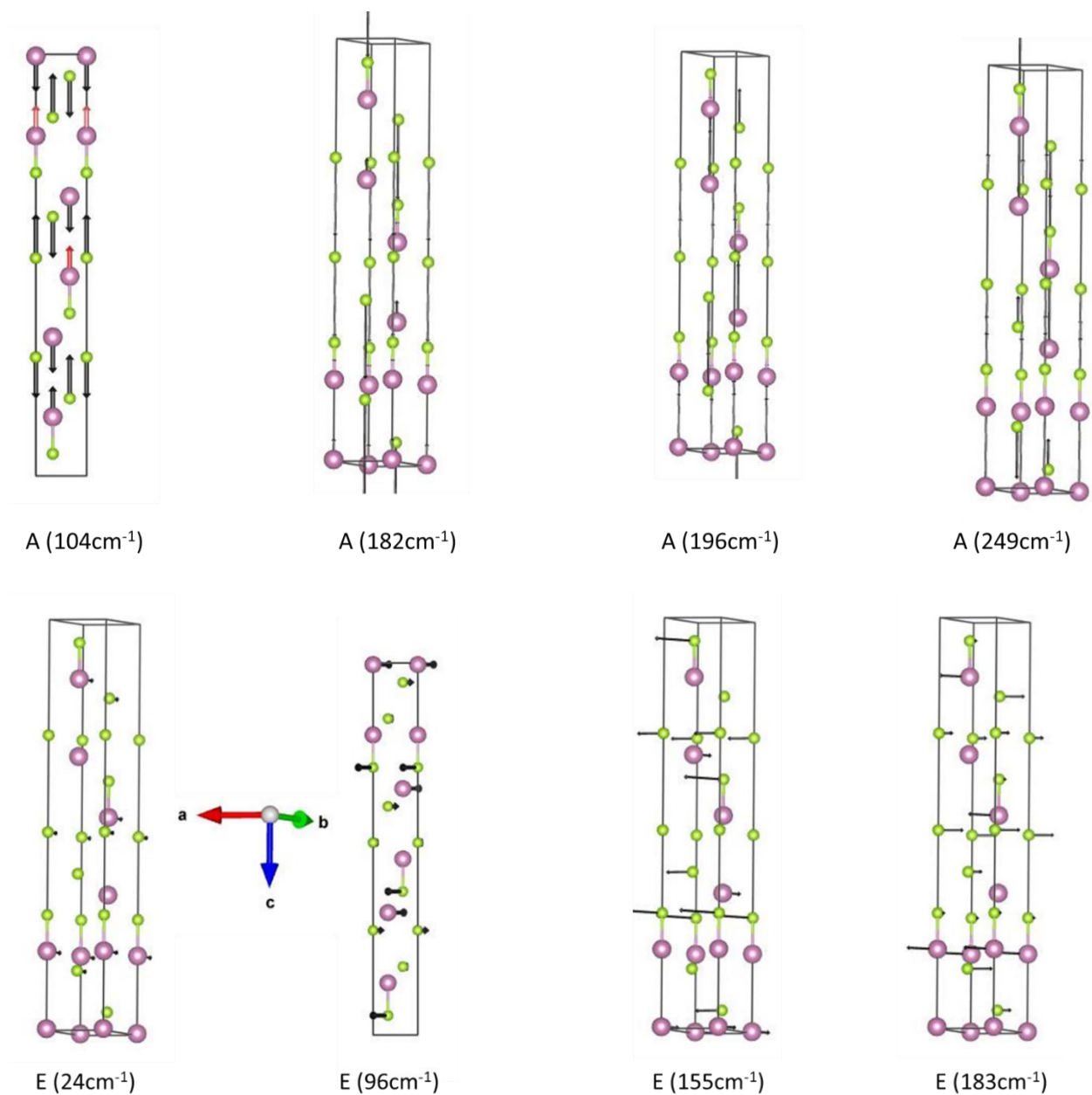

Fig S4: Theoretically calculated Eigen displacement of $A_1$ and E modes. Green and violet colour spheres represents the Se and In atoms, respectively. The black arrows represent the Eigen displacements. Length of the arrow represents the magnitude of the vibrational displacements. A1 modes are consist of out of plane atomic vibration, whereas E modes consist of in plane atomic vibration.

# Light polarization dependent photo-response study

## Note S2: Photoconductivity Experimental Setup

Photoconductivity measurements are performed in the linear two point probe configuration using Keithley 2425 source meter and Keithley 2002 multimeter. Light on/off studies are done using 1.25 mW and 1.75 mW power of 473nm and 633nm laser, respectively. Resistivity measurements was done using 1μA. Photoconductivity measurements were done along a fixed direction under different light polarization. Therefore, the electron-phonon interaction anisotropy will not take part the photoconductivity anisotropy. Only electron photon interaction anisotropy is responsible for the observed photo response anisotropy.

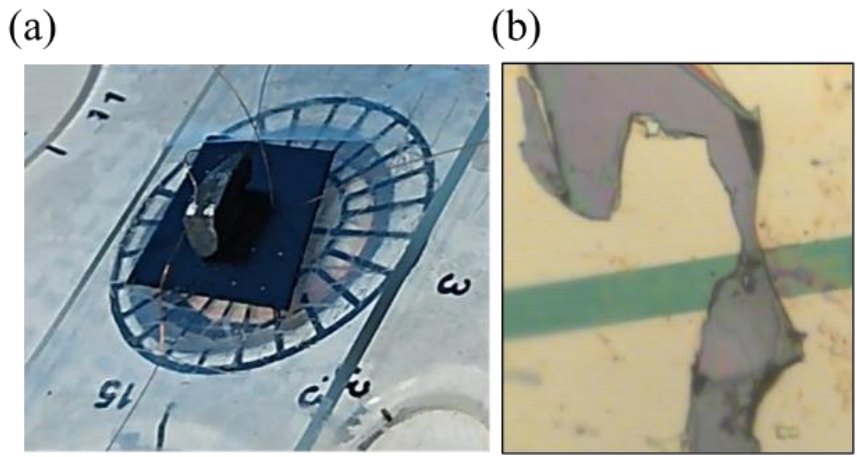

Fig S5: (a,b) Image of the polarization-dependent photoconductivity measurements geometry on *bc* plane and *ab* plane, respectively. For *bc* plane photoconductivity measurements the separation between the electrodes is 650μm, length is 5cm and hight of the crystal ~3mm. For *ab* plane photoconductivity measurements, the electrode separation 10μm, width 3.5μm and thickness 450nm.

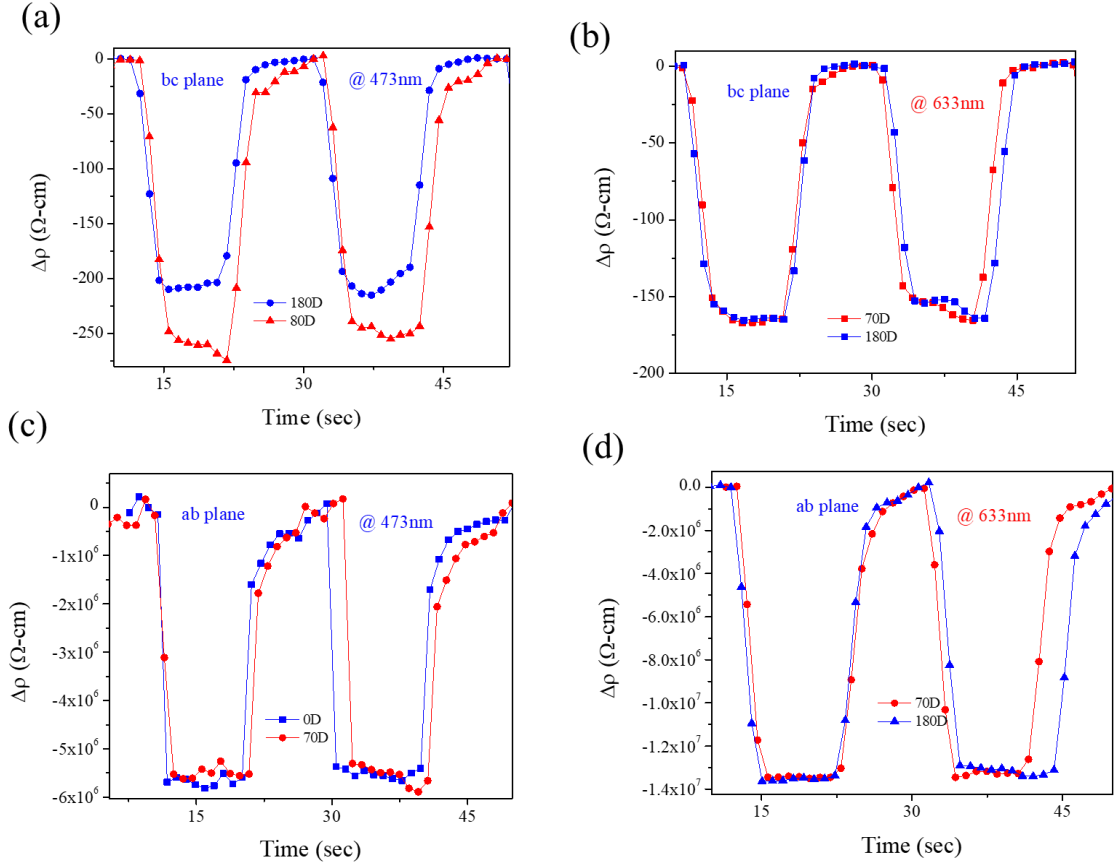

Fig S6: (a-d) Polarization-dependent Photoconductivity on *bc* plane as well as *ab* plane using both excitation sources.

**Note S3: anisotropic photo response calculation**

The photo-resistivity is calculated as $\Delta\rho = \rho_{ON} - \rho_{OFF}$. It is observed in Fig 4 that the $\Delta\rho$ is isotropic in ab plane for both the excitation laser. On the other hand the same is highly anisotropic in *bc* plane for 473 nm excitation while it is isotropic in 633 nm laser excitation. The observed photo-resistivity is maximum along *c* axis and minimum along *b* axis. The anisotropy is calculated as $\Delta\rho_{anisotropy} = \frac{\Delta\rho(c) - \Delta\rho(b)}{\Delta\rho(b)} = 15.5\%$.

**Note S4: Power dependent study**

Fig S7(c-d) show the photo resistivity under different laser power for 633 and 473 nm excitation laser, respectively. $\Delta\rho$ increase with increasing laser power and become saturated at higher powers which depicts that the observed photo response is not related to the photo thermal effect [1]. Photothermal effect (temperature increment due to photo induced heating) is increased with light power and should not saturate [2]. Noticeably, the 3R α-In2Se3 shows large photo resistivity, it is ~70% and ~35% for 633 and 473 nm laser, respectively. The resistivity under different light power is shown in Fig S7(a-b)

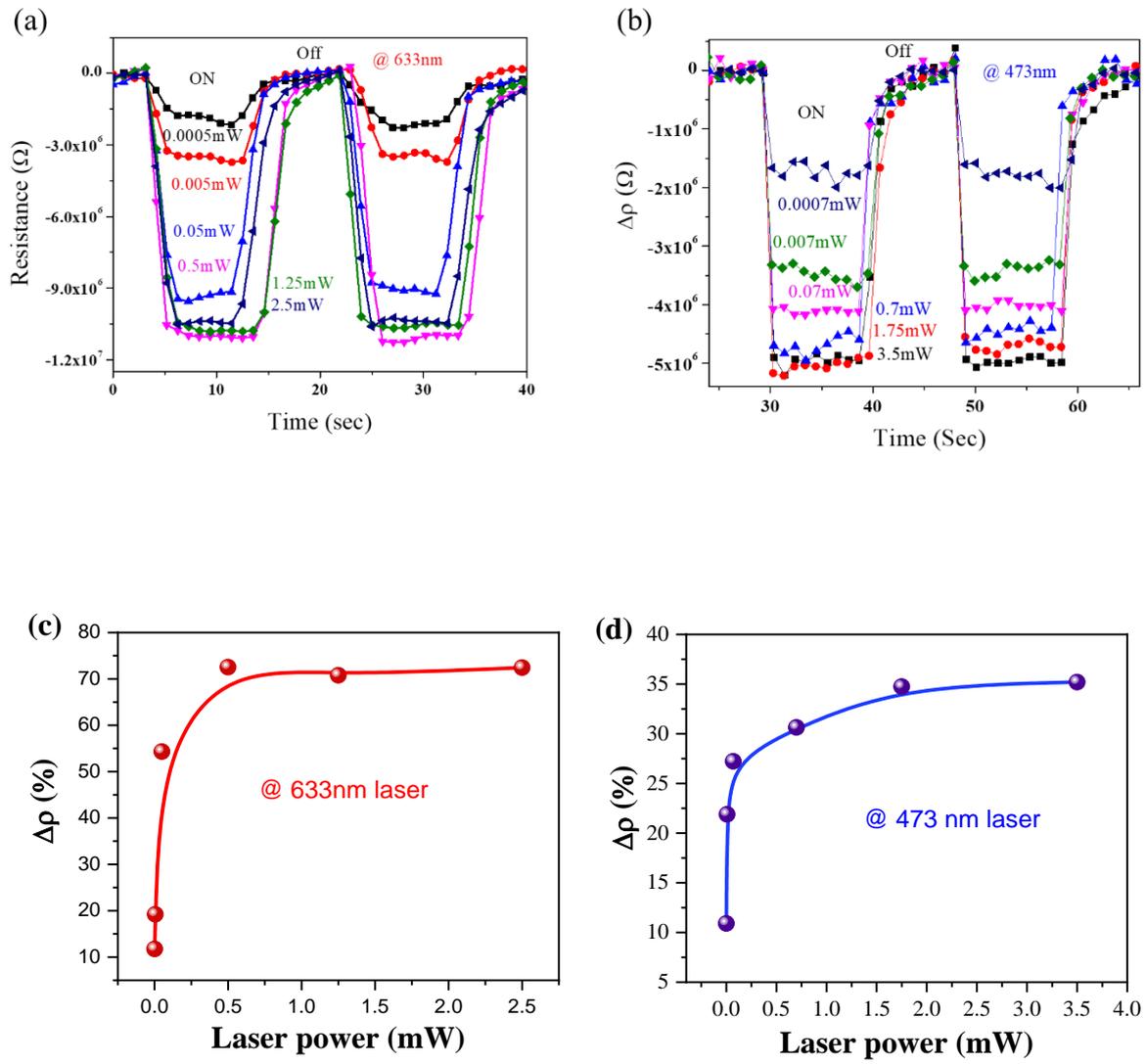

Fig S7: (a-b) resistivity changes as a function of light power. (c-d) photo resistivity under different light power.

# DFT calculated band structures

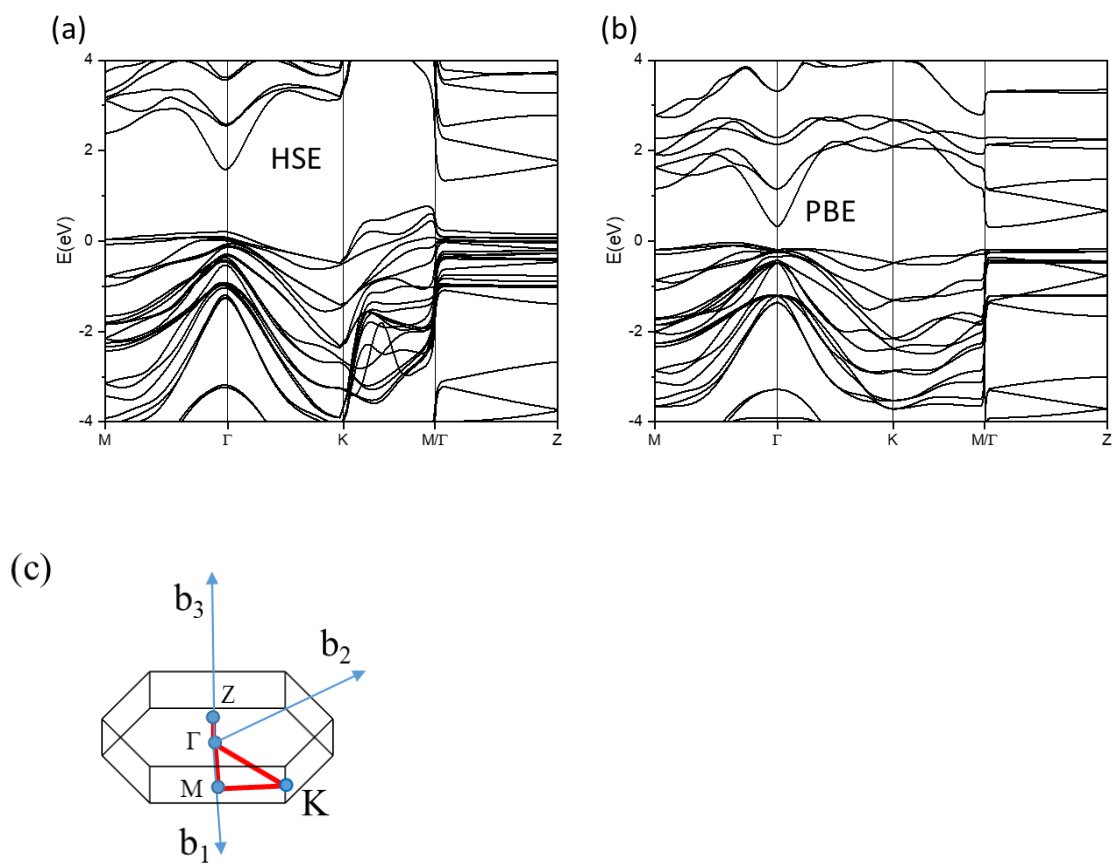

Fig S8: Band distribution along different crystallographic directions using (a) HSE06 and (b) PBE exchange correlation functional. (c) The path of high symmetry points for band structure calculation. The high symmetry k points shown in band structure plot are: $\Gamma$ (0 0 0), M (1/2 0 0), K (1/3 1/3 0) and Z (0 0 1/2). $b_1$, $b_2$ and $b_3$ are the reciprocal unit cell vectors.

# UV-VIS spectroscopy study to obtain band gap

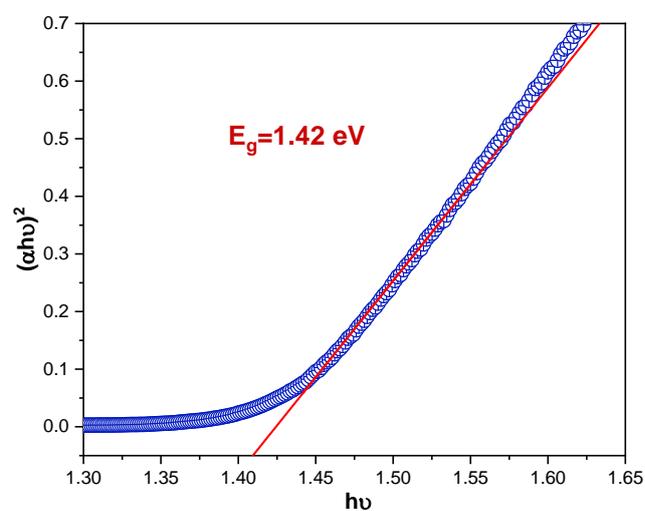

Fig S9: Experimentally observed absorption spectra, Tauc plot. Observed band gap is~1.42eV

**Note S5: V-I characteristics under different light**

Fig S10 shows the V-I characteristics of 3R α- In2Se3 under different light power. The slope of the V-I characteristics, i.e. resistivity reduce under the light power, which supports the earlier resistivity vs time results. Further, the V-I characteristics are pass through the origin indicate that the photovoltaic effect is very small [3].

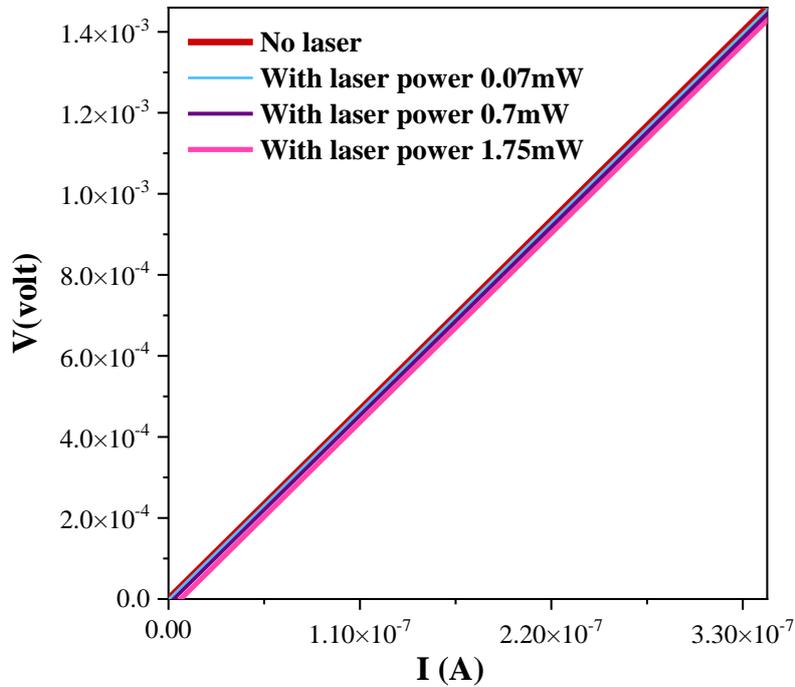

Fig S10: V–I characteristics under different light power.

**Polarization dependent reflectivity study**

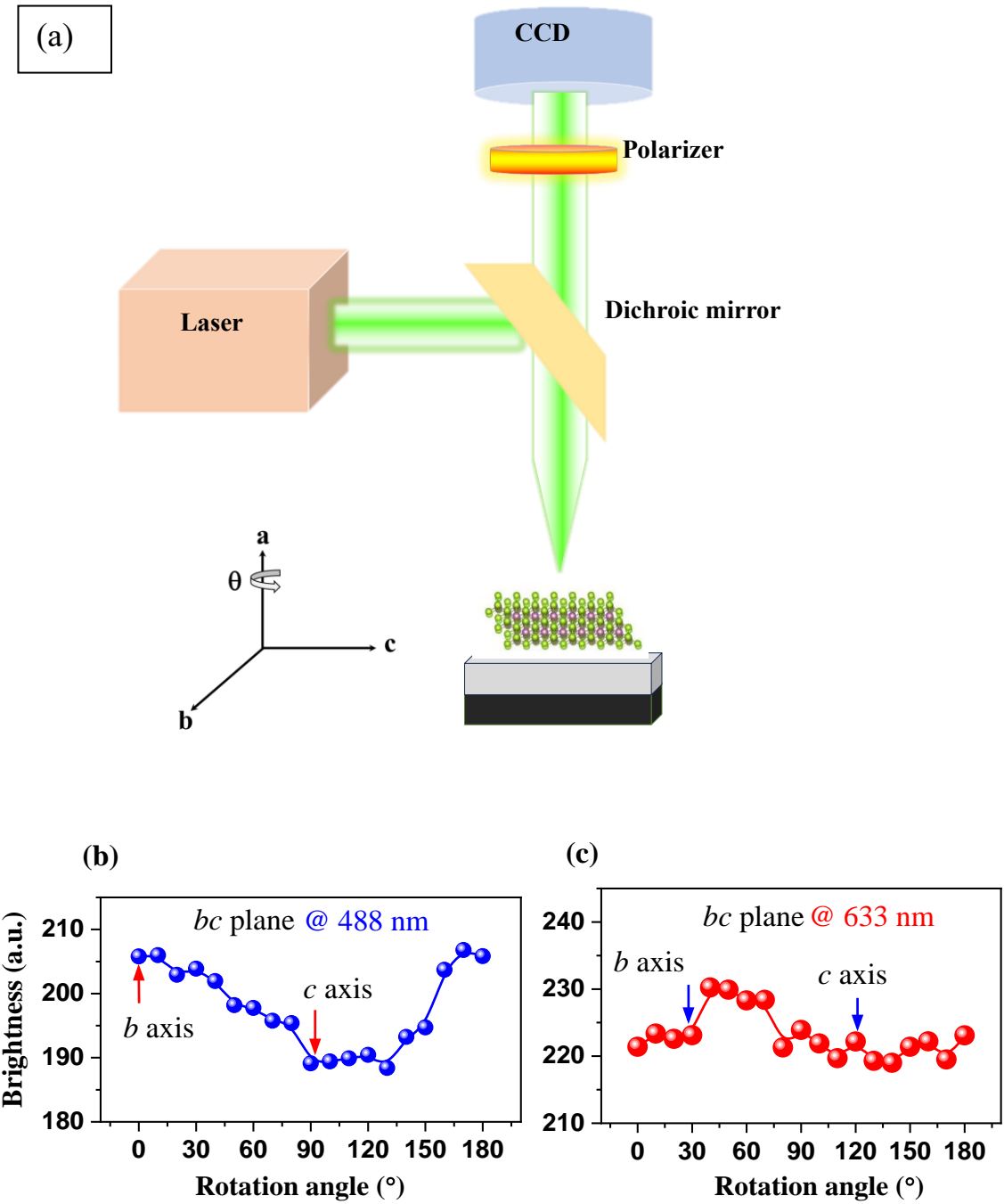

Fig S11: (a) The optical setup, configuration to capture angle-resolved reflectivity contrast images. (b) and (c) are the Relative contrast of optical reflectivity as function of light polarization angle in *bc* plane obtained using 488nm and 633 nm laser.